\documentclass[11pt, a4paper]{article}
\usepackage[text={8in,11in},margin=1.0 in,include foot]{geometry}
\usepackage{graphicx}
\usepackage{slashed}
\usepackage{amsmath}

 \title{Study of 1D stranged-charm meson family using HQET}
 \author{Pallavi Gupta, A. Upadhyay
 \\\small{\it School of Physics and Material Science},\\\small{\it Thapar University,
Patiala, Punjab-147004}\\\small{E-mail: 10gupta.pallavi@gmail.com,
alka.iisc@gmail.com}}
  
\begin{document}
\maketitle
\begin{abstract}
Recently LHCb predicted spin 1 and spin 3 states $D^{*}_{s1}(2860)$
and $D^{*}_{s3}(2860)$ which are studied through their strong
decays, and are assigned to fit the $1 ^{3}D_{1}$and $1 ^{3}D_{3}$
states in the charm spectroscopy. In this paper,using the heavy
quark effective theory, we state that assigning $D^{*}_{s1}(2860)$
as the mixing of $1^{3}D_{1}-2 ^{3}S_{1}$ states, is rather a better
justification to its observed experimental values than a pure state.
We study its decay modes variation with hadronic coupling constant
$gxh$ and the mixing angle $\theta$. We appoint spin 3 state
$D^{*}_{s3}(2860)$ as the missing 1D $3^{-} J^{P}$ state, and also
study its decay channel behavior with coupling constant $gyh$. To
appreciate the above results, we check the variation of decay modes
for their spin partners states i.e. $1D_{2}$ and $1D^{'}_{2}$ with
their masses and strong coupling constant i.e. $gxh$ and $gyh$. Our
calculation using HQET approach give mixing angle between the
$1^{3}D_{1}-2 ^{3}S_{1}$ state for $D^{*}_{s1}(2860)$ to lie in the
range (-1.6 radians$\leq \Theta\leq$-1.2 radians). Our calculation
for coupling constant values gives $gxh$ to lie between value
$0.17-0.20$ and $gyh$ to be 0.40. We expect from experiments to
observe this mixing angle to verify our results.

\noindent {\bf PACS: 14.40 Lb, 13.25 Ft}
\vspace{0.2cm}\\

\noindent{\bf Keywords: Charmed mesons ($|C|>0,B=0$), decays of
charmed mesons}
\end{abstract}
\section{Introduction}
Over the last decade many new heavy-light mesons have been observed
by various experimental collaborations. The $D^{*}_{sJ}(2860)$ was
first observed by the BaBar Collaboration in
$D_{s1}(2860)\rightarrow D^{0}K^{+},D^{+}K^{0}$ with mass
$M=2856.6\pm1.5 MeV$ and width $\Gamma = 48\pm 7 MeV$ [1]. It was
supposed to have natural parity states i.e. $
0^{+},1^{-},2^{+},3^{-}$etc. State $D_{sJ}(2860)$ as the $0^{+}$
state was ruled out after the observation of
$D_{sJ}(2860)\rightarrow
 D^{*}K$ [2]. Along with the $D^{*}K$ channel [2] BaBar also gives the
ratio R measured as $R=\frac{Br(D^{*}_{sJ}(2860)\rightarrow
D^{*}K)}{Br(D^{*}_{sJ}(2860)\rightarrow DK}=1.10$. The
$D_{sJ}(2860)$ went through extensive discussions by
 various theoretical
 models, to find a place in strange charm spectrum. Zang et al has assign the $D^{*}_{sJ}(2860)$
 as
 $2 ^{3}P_{0}$ or $1 ^{3}D_{3}$ states using the $ ^{3}P_{0}$
 model[3], Colangelo et al. assign the $D^{*}_{sJ}(2860)$ to be $1
 ^{3}D_{3}$ state using the heavy meson effective theory [4,5], D.M.Li
 et al., also favors the $D^{*}_{sJ}(2860)$ as the  $2 ^{3}P_{0}$or $1
 ^{3}D_{3}$ state using Regge phenomenology [6]. Different approaches calculated
 different value of the R ratio. Heavy Quark Effective Theory predicts R to be $\approx 0.39$[4], while $^{3}P_{0}$ model calculated it to be R = 0.59
 which is deviating from the the experimental value $R=1.10$. All
 these references favored $D^{*}_{sJ}(2860)$ as $1 ^{3}D_{s3}$ state
 due to observed narrow decay width at the cost of mismatch of R with experiments. Li and Ma assign $D_{sJ}(2860)$ to be mixing state of $1^{3}D_{1}-2 ^{3}S_{1}$ with $D^{*}_{s1}(2700)$ to be its orthogonal
 partner [7] and obtained R= 0.8 nearly close to the experimental value.

 Recently LHCb Collaboration predicted a new resonance around 2.86 GeV in the
$\overline{D}^{0} K^{-}$ invariant mass spectrum from decay channel
$B^{0}_{s}\rightarrow \overline{D}^{0}K^{-}\Pi^{+}$, containing the
mixture of spin-1 and spin-3 states components corresponding to
$D^{*}_{s1}(2860)$ and $D^{*}_{s3}(2860)$ [8,9] where the mass and
width parameters are
\begin{center}
$M(D^{*}_{s1}(2860)= 2859\pm 12\pm6\pm23 MeV$

 $\Gamma(D^{*}_{s1}(2860)=159\pm23\pm27\pm72 MeV $

 $M(D^{*}_{s3}(2860)=2860.5\pm2.6 \pm2.5\pm6.0MeV$

 $\Gamma(D^{*}_{s3}(2860)=53\pm 7\pm4\pm6 MeV $

\end{center}
Here the first error is statistical error, second is the
experimental systematic effects and the last one is due to model
variations. LHCb observed two $D^{*}_{s}(2860)$ states with spin 1
and spin 3. From the previous study it can be speculated that it is
a spin 3 resonance of $D^{*}_{s}(2860)$ that belongs to
$1^{3}D_{s3}$ state, with a narrow width $\Gamma=53 MeV$.
Theoretically, R value can be matched with the experimental value,
considering its contribution coming from spin 1 $D^{*}_{s1}(2860)$
resonance. Comparing the earlier theoretical mass predictions,
$D^{*}_{s1}(2860)$ can be assumed to fit in $1 ^{3}D_{s1}$ state of
1D family or can be a mixture of $1^{3}D_{1}$ and $2 ^{3}S_{1}$
states. Assigning $D^{*}_{s1}$ as a mixing state of
$1^{3}D_{1}-2^{3}S_{1}$ may be a better justification than assigning
it as a pure state, because the R value calculated here matches with
the experimental data. By choosing suitable mixing angle $(\theta)$,
the calculated R value is better justified with the experimental
value.  X.H.Zhong by chiral quark model [10,11] studied the
$D^{*}_{sJ}$ state as the $1 ^{3}D_{3}$ state with some $1
^{3}D_{2}-1 ^{1}D_{2}$ mixing. Wang [12] tried to reproduce the
experimental value R= 1.10 with some suitable hadronic coupling
constants, by including chiral symmetry breaking corrections in
heavy quark effective theory. Besides these studies, Vijande et al.,
also assign $D_{sJ}(2860)$ to be the multi-quark exotic state as $
c\overline{s}-cn\overline{s}$ $\overline{n}$ [13]. Stephen Godfrey
by adopting the pseudoscalar emission decay model[14], Qing-Tao Song
by QPC model [15] studied $D^{*}_{sJ}$ as $1^{3}D_{1}-2^{3}S_{1}$.
Various predictions are made to study the mixing effects in
$D^{*}_{sJ}$ state [16,17,18,19].

In Particle data Group [20] 1S and 1P stranged charm states are
nicely
 described, but information for other states is still missing. Thus
 $D^{*}_{s1}(2860)$ and $D^{*}_{s3}(2860)$ can be fitted in 1D state
 family. The strange meson states with $J^{P}$ states predicted by various theoretical model is gathered in Table\ref{tab1}
\begin{table}[h]
\begin{center}

\begin{tabular}{|c|c|c|c|c|}
  \hline
   & GI[21] & PE[22] & EFG[23] & Expt.[20] \\
  $J^{P}(^{2s+1}L_{j})$&(MeV)&(MeV)&(MeV)&(MeV)\\
  \hline
 $ 0^{-}(^{1}S_{0})$ & 1979 & 1965 & 1969& 1968 \\
$1^{-}(^{3}S_{1}) $& 2129 & 2113 & 2111 & 2112 \\
 $ 0^{+}(^{3}P_{0})$ & 2484 & 2487 & 2509 & 2318 \\
 $ 1^{+}(^{1}P_{1})$& 2459 & 2535 & 2536 & 2460 \\
$  1^{+}(^{3}P_{1})$ & 2556 & 2605& 2574 & 2536 \\
 $ 2^{+}(^{3}P_{2})$ & 2592 & 2581 & 2571 & 2573 \\
 $ 1^{-}(^{3}D_{1})$ & 2899 & 2913 & 2913 & 2859 \\
 $  2^{-}(^{2}D_{2})$ & 2900 & 2900 & 2931 & - \\
 $  2^{-}(^{3}D_{2}^{,})$ & 2926 & 2953 & 2961 & - \\
  $ 3^{-}(^{3}D_{3})$ & 2917& 2925 & 2871 & 2860 \\
  \hline
\end{tabular}
\caption{Theoretically predicted masses}\label{tab1}

\end{center}
\end{table}

 Table\ref{tab1} shows the state and decay modes of 1D family. Here it is seen, that complete information about the states and decay modes is still missing, thus more theoretical and experimental efforts are to be
 made. In this paper, we study the strong decays and R values
 for 1D family using heavy quark effective approach and analyze the mixing effects in such states. In the past years, HQET has been successful in assigning suitable
$J^{P}$ states to the observed D and B-mesons using their decays
widths in terms of coupling constants. We use HQET approach to study
$D^{*}_{s1}(2860)$
 as a pure state, and also as a mixture of $1^{3}D_{1}$ and $2 ^{3}S_{1}$
 states and would rather classify them on the basis of their $J^{P}$ state analyzing their strong decays channels and hadronic
 strong coupling constants $gxh$ and $gyh$.
 $D^{*}_{s3}(2860)$ is assigned the $1 ^{3}D_{3}$ position in 1D strange charm
 mesons. To complete the 1D  family, we also try to study the behavior of their spin partners
 i.e. $1D_{2}$ and $1D_{2'}$ which are still missing experimentally.

 The paper is divided in the following sections. Section 2 describes the heavy quark effective theory formalism used for the strong decays. Section 3 discusses the
 members of 1D family. In this section, all
the four states with their decay modes in terms of their couplings
are described in different subsections. To appreciate the
experimental value of R, various mixing effects in terms of mixing
angle theta is studied. We finally conclude our results in section
4.

 \section{Framework}
In the heavy quark limit $m_{Q}>> \Lambda_{QCD}>>m_{q}$,
$Q\overline{q}$ system can be effectively studied using Heavy quark
effective theory. According to this theory, heavy quark acts like
static color source with spin $s_{Q}$, which due to heavy flavor
symmetry, interacts only with the light degree of freedom having
spin $s_{l}$ through the exchange of soft gluons. This picture can
be compared with that of hydrogen atom [24]. The basic idea is that
in a $Q\overline{q}$ system, heavy quark plays the role of a nucleus
and the light quark plays the role of an electron. This
$Q\overline{q}$ system can be categorized in doublets in relation to
the total conserved angular momentum $i.e.$
$s_{l}=s_{\overline{q}}+L$, where $s_{\overline{q}}$ and L are the
spin and orbital angular momentum
 of the light anti-quark respectively. For L=0(S-wave) the
doublet is represented by $(D,D^{*})$ with $J^{P}_{s_{l}}=
(0^{-},1^{-})_{\frac{1}{2}}$, which for L=1(P-wave) , there are two
doublets represented by $(D^{*}_{0},D_{1})$ and
$(D_{1}^{'},D^{*}_{2})$ with
$J^{P}_{s_{l}}=(0^{+},1^{+})_{\frac{1}{2}}$ and
$(1^{+},2^{+})_{\frac{3}{2}}$ respectively. Two doublets of L=2
(D-wave) are represented by $(D^{*}_{1},D_{2})$ and
$(D_{2}^{'},D^{*}_{3})$ belonging to
$J^{P}_{s_{l}}=(1^{-},2^{-})_{\frac{3}{2}}$ and
$(2^{-},3^{-})_{\frac{5}{2}}$ respectively. These doublets are
described by the effective super-field
$H_{a},S_{a},T_{a},X_{a},Y_{a}$ [25], where the field $H_{a}$
describe the $(D,D^{*})$ doublet i.e. S-wave, $S_{a}$ and $T_{a}$
fields represents the P-wave doublets $(0^{+},1^{+})_{\frac{1}{2}}$
and $(1^{+},2^{+})_{\frac{3}{2}}$ respectively. D-wave doublets are
represented by the $X_{a}$ and $Y_{a}$ fields. These fields are as
\begin{gather}
\label{eq:lagrangian}
 H_{a}=\frac{1+\slashed
v}{2}\{P^{*}_{a\mu}\gamma^{\mu}-P_{a}\gamma_{5}\}\\
S_{a}=\frac{1+\slashed
v}{2}\{P^{\mu}_{1a}\gamma_{\mu}\gamma_{5}-P^{*}_{0a}\}\\
T^{\mu}_{a}=\frac{1+\slashed v}{2}
\{P^{*\mu\nu}_{2a}\gamma_{\nu}-P_{1a\nu}\sqrt{\frac{3}{2}}\gamma_{5}
[g^{\mu\nu}-\frac{\gamma^{\nu}(\gamma^{\mu}-\upsilon^{\mu})}{3}]\}\\
X^{\mu}_{a}=\frac{1+\slashed
v}{2}\{P^{\mu\nu}_{2a}\gamma_{5}\gamma_{\nu}-P^{*}_{1a\nu}\sqrt{\frac{3}{2}}[g^{\mu\nu}-\frac{\gamma^{\nu}(\gamma^{\mu}+\gamma^{\mu})}{3}]\}\\
Y^{\mu\nu}_{a}=\frac{1+\slashed
v}{2}\{P^{*\mu\nu\sigma}_{3a}\gamma_{\sigma}-P^{\alpha\beta}_{2a}\sqrt{\frac{5}{3}}\gamma_{5}[g^{\mu}_{\alpha}g^{\nu}_{\beta}-
\frac{g^{\nu}_{\beta}\gamma_{\alpha}(\gamma^{\mu}-v^{\mu})}{5}-\frac{g^{\mu}_{\alpha}\gamma_{\beta}(\gamma^{\nu}-v^{\nu})}{5}]\}
\end{gather}

The light pseudoscalar mesons are described by the fields $\xi=
exp^{\frac{iM}{f}}$. The pion octet is introduced by the vector and
axial combinations
$V^{\mu}=\frac{1}{2}\xi\partial^{\mu}\xi^{\dag}+\xi^{\dag}\partial^{\mu}\xi$
and
$A^{\mu}=\frac{1}{2}\xi\partial^{\mu}\xi^{\dag}-\xi^{\dag}\partial^{\mu}\xi$.
We choose $f_{\pi}=130MeV$. Here, all traces are taken over Dirac
spinor indices, light quark $SU(3)_{V}$ flavor indices a = u, d, s
and heavy quark flavor indices Q = c, b [25]. The Dirac structure of
chiral Lagrangian has been replaced by velocity vector v. At the
leading approximation, the heavy meson chiral lagrangian $L_{HH}$,
$L_{SH}$, $L_{TH}$, $L_{XH}$, $L_{XS}$, $L_{XT}$, $L_{YH}$,
$L_{YS}$, $L_{YT}$ for
 the two-body strong decays to light pseudoscalar mesons can be
 written as :
\begin{center}
\begin{gather}
\label{eq:lagrangian}
 L_{HH}=g_{HH}Tr\{\overline{H}_{a}
 H_{b}\gamma_{\mu}\gamma_{5}A^{\mu}_{ba}\}\\
L_{SH}=g_{SH}Tr\{\overline{H}_{a}S_{b}\gamma_{\mu}\gamma_{5}A^{\mu}_{ba}\}+h.c.\\
L_{TH}=\frac{g_{TH}}{\Lambda}Tr\{\overline{H}_{a}T^{\mu}_{b}(iD_{\mu}\slashed
A + i\slashed D A_{\mu})_{ba}\gamma_{5}\}+h.c.\\
L_{XH}=\frac{g_{XH}}{\Lambda}Tr\{\overline{H}_{a}X^{\mu}_{b}(iD_{\mu}\slashed
A + i\slashed D A_{\mu})_{ba}\gamma_{5}\}+h.c.\\
L_{XS}=\frac{g_{XS}}{\Lambda}Tr\{\overline{S}_{a}X^{\mu}_{b}(iD_{\mu}\slashed
A + i\slashed D A_{\mu})_{ba}\gamma_{5}\}+h.c.\\
L_{XT}=\frac{1}{\Lambda^{2}}Tr\{\overline{T}^{\mu}_{a}X^{\nu}_{b}[k^{T}_{1}\{D_{\mu}
,D_{\nu}\}A_{\lambda}+k^{T}_{2}(D_{\mu}D_{\lambda}A_{\nu}+D_{\nu}D_{\lambda}A_{\mu})]_{ba}\gamma^{\lambda}\gamma_{5}\}+h.c.\\
L_{YH}=\frac{1}{\Lambda^{2}}Tr\{\overline{H}_{a}Y^{\mu\nu}_{b}[k^{H}_{1}\{D_{\mu}
,D_{\nu}\}A_{\lambda}+k^{H}_{2}(D_{\mu}D_{\lambda}A_{\nu}+D_{\nu}D_{\lambda}A_{\mu})]_{ba}\gamma^{\lambda}\gamma_{5}\}+h.c.
\\L_{YS}=\frac{1}{\Lambda^{2}}Tr\{\overline{S}_{a}Y^{\mu\nu}_{b}[k^{S}_{1}\{D_{\mu}
,D_{\nu}\}A_{\lambda}+k^{S}_{2}(D_{\mu}D_{\lambda}A_{\nu}+D_{\nu}D_{\lambda}A_{\mu})]_{ba}\gamma^{\lambda}\gamma_{5}\}+h.c.
\\L_{YT}=\frac{g_{YT}}{\Lambda}Tr\{\overline{T}_{a\mu}X^{\mu\nu}_{b}(iD_{\nu}\slashed
A + i\slashed D A_{\nu})_{ba}\gamma_{5}\}+h.c.
%\end{split}
%\end{equation}
\end{gather}
%\end{align*}
\end{center}
 From the chiral
lagrangian terms $L_{HH},L_{SH},L_{TH},L_{XH},L_{YH}$, the two body
strong decay of $Q\overline{q}$ system to final state light
pseudo-scalar mesons M $(\Pi,\eta,K)$ can be described as

$(1^{-},2^{-}) \rightarrow (0^{-},1^{-}) + M$
\begin{gather}
\label{eq:lagrangian} \Gamma(1^{-} \rightarrow 0^{-})=
C_{M}\frac{4g_{X}^{2}}{9\Pi f_{\Pi}^{2}\Lambda^{2}}
\frac{M_{f}}{M_{i}}[p_{M}^{3}(m_{M}^{2}+p_{M}^{2})]\\
\Gamma(1^{-} \rightarrow 1^{-})= C_{M}\frac{2g_{X}^{2}}{9\Pi
f_{\Pi}^{2}\Lambda^{2}}
\frac{M_{f}}{M_{i}}[p_{M}^{3}(m_{M}^{2}+p_{M}^{2})]\\
\Gamma(2^{-} \rightarrow 1^{-})= C_{M}\frac{2g_{X}^{2}}{3\Pi
f_{\Pi}^{2}\Lambda^{2}}
\frac{M_{f}}{M_{i}}[p_{M}^{3}(m_{M}^{2}+p_{M}^{2})]
\end{gather}

$(2^{-},3^{-}) \rightarrow (0^{-},1^{-}) + M$

\begin{gather}
\label{eq:lagrangian} \Gamma(2^{-} \rightarrow 1^{-})=
C_{M}\frac{4g_{Y}^{2}}{15\Pi f_{\Pi}^{2}\Lambda^{4}}
\frac{M_{f}}{M_{i}}[p_{M}^{7}]\\
\Gamma(3^{-} \rightarrow 0^{-})= C_{M}\frac{4g_{Y}^{2}}{35\Pi
f_{\Pi}^{2}\Lambda^{4}} \frac{M_{f}}{M_{i}}[p_{M}^{7}]\\
\Gamma(3^{-} \rightarrow 1^{-})= C_{M}\frac{16g_{Y}^{2}}{105\Pi
f_{\Pi}^{2}\Lambda^{4}} \frac{M_{f}}{M_{i}}[p_{M}^{7}]
\end{gather}

$(1^{-},2^{-}) \rightarrow (0^{+},1^{+}) + M$

\begin{gather}
\label{eq:lagrangian} \Gamma(2^{-} \rightarrow 1^{+})=
C_{M}\frac{2g_{X}^{2}}{5\Pi f_{\Pi}^{2}\Lambda^{2}}
\frac{M_{f}}{M_{i}}[p_{M}^{5}]\\
\Gamma(2^{-} \rightarrow 0^{+})= C_{M}\frac{4g_{X}^{2}}{15\Pi
f_{\Pi}^{2}\Lambda^{2}} \frac{M_{f}}{M_{i}}[p_{M}^{5}]\\
\Gamma(1^{-} \rightarrow 1^{+})= C_{M}\frac{2g_{X}^{2}}{3\Pi
f_{\Pi}^{2}\Lambda^{2}} \frac{M_{f}}{M_{i}}[p_{M}^{5}]
\end{gather}
$(2^{-},3^{-}) \rightarrow (0^{+},1^{+}) + M$
\begin{gather}
\label{eq:lagrangian} \Gamma(3^{-} \rightarrow 1^{+})=
C_{M}\frac{4g_{Y}^{2}}{15\Pi f_{\Pi}^{2}\Lambda^{4}}
\frac{M_{f}}{M_{i}}[p_{M}^{5}(m_{M}^{2}+p_{M}^{2})]\\
\Gamma(2^{-} \rightarrow 1^{+})= C_{M}\frac{8g_{Y}^{2}}{75\Pi
f_{\Pi}^{2}\Lambda^{4}}
\frac{M_{f}}{M_{i}}[p_{M}^{5}(m_{M}^{2}+p_{M}^{2})]\\
\Gamma(2^{-} \rightarrow 0^{+})= C_{M}\frac{4g_{Y}^{2}}{25\Pi
f_{\Pi}^{2}\Lambda^{4}}
\frac{M_{f}}{M_{i}}[p_{M}^{5}(m_{M}^{2}+p_{M}^{2})]
\end{gather}
In the above expressions of decay width,  $M_{i},M_{f}$ stands for
initial and final meson mass, $g_{X}$ and $g_{Y}$ are hadronic
coupling constants, $\Lambda$ is the chiral symmetry breaking scale
= 1GeV, $p_{M}$ and $m_{M}$ is the final momentum and mass of the
emitted light pseudo-scalar meson. The coefficient
$C_{\Pi^{\pm}},C_{K^{\pm}},C_{K^{0}},C_{\overline{K}^{0}}=1$,$C_{\Pi^{0}}=\frac{1}{2}$
and $C_{\eta}=\frac{2}{3} or \frac{1}{6}$ [25]. Different values of
$C_{\eta}$ corresponds to the initial state being
$c\overline{u},c\overline{d}$ or $ c\overline{s}$ respectively.
 \section{Numerical Results}

  OZI allowed two-body strong decays of 1D strange charm family are calculated using the
  heavy quark effective approach given in section 2. Partial and total decay widths of
  these states are studied and compared with the experimental
  values. OZI allowed decay channels for $D^{*}_{s1}(2860)$ and $D^{*}_{s3}$ are $DK$, $D^{*}K$, $D_{s}\eta$ and $D^{*}_{s}\eta$,
  and for their spin partners $1D_{s2}$ and $1D^{'}_{s2}$ decay modes are $D^{*}K$, $D^{*}_{s}\eta$, $D(2400)K$ and $D^{*}_{s}(2317)\eta$. For this, we take input parameters as the initial masses for
$D^{*}_{s1}$ and $D^{*}_{s3}$, given by the
 LHCb [8,9] and 2890 MeV and 2900 MeV for their spin partner states $1D_{s2}$ and $1D^{'}_{s2}$ respectively. Heavy Quark Effective Theory shows that, decay widths
 also depend on the strong hadronic coupling  $g_{XH}$, $g_{YH}$, $g_{XS}$ and
 $g_{YS}$. The strong couplings have been constrained to be with 0
 and 1 [26] but their experimental information is still missing. In the next subsections, we calculated two of these coupling
constants i.e. $g_{XH}$ and $g_{YH}$ using the decay
 widths and other available experimental data.

 \subsection{$D^{*}_{s1}(2860)$}

 $D^{*}_{s1}(2860)$ was first observed by BaBar collaboration and in 2014 its spin, mass and decay width was confirmed by
 LHCb. In this subsection, heavy quark effective theory is adopted to reproduce the experimental data given by these
 collaborations. The coupling constant $gxh$ is obtained and $D^{*}_{s1}(2860)$ state are assigned as the $1^{-}$ member of the 1D charm family. Assuming it to be the pure $1D$ $1^{-}$ state, we calculated the
total and partial decay widths of $D^{*}_{s1}(2860)$ using the
  decay width formulae given in section 2 in terms of their hadronic coupling
  constants. These partial decay widths and ratios are tabulated in
  Table\ref{tab2}. Along with the partial decay widths , we also studied the ratios such as

\begin{gather}
\label{eq:lagrangian}
 R=\frac{Br(D^{*}_{s1}(2860)\rightarrow
D^{*}K)}{Br(D^{*}_{s1}(2860)\rightarrow DK}\\
R1=\frac{Br(D^{*}_{s1}(2860)\rightarrow
D_{s}\eta)}{Br(D^{*}_{s1}(2860)\rightarrow DK}
\end{gather}

  \begin{table}[h]
  \begin{center}
   \begin{tabular}{lccccccr}
   \hline
   \hline

Theory&$DK$&$D^{*}K$&$D_{s}\eta$&$D^{*}_{s}\eta$&Total&$\frac{D^{*}K}{DK}$&$\frac{D_{s}\eta}{DK}$\\
\hline
 Our &2865.45$gxh^{2}$&693.135$gxh^{2}$&508.189$gxh^{2}$&85.70$gxh^{2}$&4152.48$gxh^{2}$&0.24&0.177\\
 \hline
 Experimental[8]&&&&&159&1.10&\\

 \hline
    \hline
   \end{tabular}
    \end{center}
\caption{Calculated partial and total decay widths of
$D^{*}_{s1}(2860)$ as pure $(1^{3}D_{1})$}\label{tab2}

\end{table}

It can be seen from the Table\ref{tab2} that our calculated R value
does not matches with the experimental value 1.10. Same has also
been calculated by various theoretical models like Ref[15] gives
$\Gamma(\frac{D^{*}K}{DK})= 0.46 to 0.70$ and
$\Gamma(\frac{D_{s}\eta}{DK})= 0.10 to 0.14$,Ref[18] gives
$\Gamma(\frac{D^{*}K}{DK})= 12.5 to 7.6$ and
$\Gamma(\frac{D_{s}\eta}{DK})= 0.30 to 0.14$ and Ref[5] gives
$\Gamma(\frac{D^{*}K}{DK})= 0.06$ and $\Gamma(\frac{D_{s}\eta}{DK})=
 0.23$. As it can be seen that R value i.e. $\Gamma(\frac{D^{*}K}{DK})$
calculated by our HQET approach and by other theoretical approaches
[15,18,5] does not matches with the experimental R value $i.e.$
1.10. As R is independent of couplings, so to justify the
experimental value of R, we include the mixing of the states. Here
we constraint the hadronic coupling constant $gxh$ from the
literature. According to this scheme, state $D^{*}_{s1}(2860)$ is
assumed to be the  mixture of $2^{3}S_{1}$ and $1^{3}D_{1}$ states
with $D_{s}(2700)$ to be its  orthogonal partner satisfying the
relation

 \begin{center}
$ \left(
\begin{array}{c}
  D_{s1}(2S) \\
  D_{s1}(2860) \\
\end{array}
\right) = \left(
\begin{array}{cc}
Cos\theta & Sin\theta \\
  -Sin\theta & Cos\theta \\
\end{array}
\right) \left(
\begin{array}{c}
   2^{3}S_{1} \\
  1^{3}D_{1} \\
\end{array}
\right) $
 \end{center}

where $\theta$ is the mixing angle. Effect of variation of total
decay width of $D_{s1}(2860)$ state with coupling constant for
different mixing angle is shown in Figure1 to Figure4. We have seen
this variation for some typical values of mixing angle at $\theta=
0^{\circ}$, $\theta= -30^{\circ}$ and for $\theta= -60^{\circ}$ and
$\theta= -80^{\circ}$ where $\theta= 0^{\circ}$ correspond to non
mixing i.e. pure $1^{3}D_{1}$.

%\begin{figure}[h!]

 % \centering
  %  \includegraphics[width=2in]{d1n0d}
%\includegraphics[width=2in]{d1n15d}
%\includegraphics[width=2in]{d1n30d}
 %    \caption{$1(a)$ shows decay channels for $\theta = 0$ , $1(b)$
%shows for $\theta = 15$ and $1(c)$ shows for $\theta = 30$.}
%\end{figure}
\begin{figure}[ht]
  \begin{minipage}[b]{0.5\linewidth}
   % \includegraphics[width=.7\linewidth]{d1n0d}
    %\caption{Decay widths of $D_{s1}(2860)$ for $\theta = 0^{\circ}$}
    \caption{\small{Decay widths of $D_{s1}(2860)$ for $\theta = 0^{\circ}$}}
    \label{ fig1:a)}
  \end{minipage}
  \begin{minipage}[b]{0.5\linewidth}
    \caption{\small{Decay widths of $D_{s1}(2860)$ for$ \theta =
    -30^{\circ}$}}
    \label{ fig1:b}
  \end{minipage}
\begin{minipage}[b]{0.5\linewidth}
    \caption{\small{Decay widths of $D_{s1}(2860)$ for $\theta =
    -60^{\circ}$}}
    \label{ fig1:a)}
  \end{minipage}
  \begin{minipage}[b]{0.5\linewidth}
    \caption{\small{Decay widths of $D_{s1}(2860)$ for $\theta =
    -80^{\circ}$}}
    \label{ fig1:b}
  \end{minipage}
  \end{figure}

 Figure1 - Figure4 shows that $DK$ is the main decay channel of
 this state. Apart from $DK$, $D^{*}K$ and $D_{s}\eta$ are also
 important decay channels of $1^{3}D_{1}$, whereas the calculated decay
 width for $D^{*}_{s}\eta$ is found to be small. Dominance of $D^{*}_{s}\eta$ decay channel enhances with more mixing effect. R ratio defined in the section 1 now depends on the mixing angle and
strong coupling constants $gxh$ and $ghh$. Fixing $ghh=0.17$ [25]
variation of R value with the mixing angle is seen in figure 5. This
figure shows that experimental R value 1.10 can be achieved
theoretically corresponding to the mixing angle $ -1.6 \leq \theta
\leq -1.2$ radians. For this range of mixing angle our hadronic
coupling constant comes out to be $ 0.17 \leq gxh \leq 0.20$. This
variation of $gxh$ can be shown in Figure 6.

%\begin{figure}[h!]

 % \centering
%\includegraphics[width = 2.2in]{d1nra}
%\includegraphics[width = 2.2in]{d1nga}
%\caption{$2(a)$ shows the variation of R value with mixing angle and
%$2(b)$ shows the variation for coupling constant $gxh$ and mixing
%angle.}
%\end{figure}
\begin{figure}[ht]
  \begin{minipage}[b]{0.5\linewidth}
    \caption{\small{Variation of R value with mixing angle}}
        \label{ fig1:a)}
  \end{minipage}
  \begin{minipage}[b]{0.5\linewidth}
    \caption{\small{Variation of cupling constant gxh with mixing
    angle}}
    \label{ fig1:b}
  \end{minipage}
  \end{figure}

 For these calculated values of mixing angle and coupling
constant, partial and total decay width are again studied. Total
width $\Gamma$ comes out to be 159 MeV, which matches very well with
the experimental. Other partial
  decay widths are listed in Table\ref{tab3}. R value shown in column 7 comes
  out to be 0.91 which now is close to the the experimental observed
  value.

 \begin{table}[h]
  \begin{center}
   \begin{tabular}{lccccccr}
   \hline
   \hline

Theory&$DK$&$D^{*}K$&$D_{s}\eta$&$D^{*}_{s}\eta$&Total&$\frac{D^{*}K}{DK}$&$\frac{D_{s}\eta}{DK}$\\
&(MeV)&(MeV)&(MeV)&(MeV)&(MeV)&&\\
 \hline
 Our &60.61&55.26&11.85&18.66&146.39&0.91&0.19\\
 \hline
 Experimental&&&&&159&1.10&\\
 \hline
    \hline
   \end{tabular}
\caption{Calculated partial and total decay widths of
$D^{*}_{s1}(2860)$ as a mixture of $(1^{3}D_{1})$ and
$2^{3}S_{1}$}\label{tab3}
   \end{center}
\end{table}

\subsection{$D^{*}_{s3}(2860)$}

Considering $D^{*}_{s3}(2860)$ as the $1^{3}D_{s3}$, its decay
channels and partial decay widths are presented in Table\ref{tab4}.
Decay widths are listed in terms of hadronic coupling constant
$gyh$. Figure7 shows the variation of the partial and total decay
width with this coupling constant.

\begin{table}[h]
  \begin{center}
   \begin{tabular}{lccccccr}
   \hline
   \hline

Theory&$DK$&$D^{*}K$&$D_{s}\eta$&$D^{*}_{s}\eta$&Total\\
&(MeV)&(MeV)&(MeV)&(MeV)&(MeV)&\\
 \hline
 Our&249.18$gyh^{2}$&96.39$gyh^{2}$&24.72$gyh^{2}$&4.54$gyh^{2}$&374.846$gyh^{2}$\\
 \hline
 Experimental&&&&&53\\
 \hline
    \hline
   \end{tabular}
\caption{Calculated partial and total decay widths of
$D^{*}_{s3}(2860)$ as $1^{3}D_{s3}$}\label{tab4}
   \end{center}
\end{table}

\begin{figure}[h!]

  \centering
\caption{shows the variation of partial decay widths of
$D^{*}_{s3}(2860)$ as the $1^{3}D_{s3}$ state with hadronic
coupling.}
\end{figure}

 This picture clearly shows that $DK$ is the dominant decay
mode of $D^{*}_{s3}(2860)$. Other important decay channels are
$D^{*}K$, $D_{s}\eta$ with $D^{*}_{s}\eta$ contributing least.
Computing it with the experimental value of total decay width
$\Gamma=53 MeV$, coupling constant $gyh$ comes out to be 0.40. These
partial decay widths can be used to calculate the ratio R.
\begin{gather}
\label{eq:lagrangian} R= \frac{\Gamma(D^{*}_{s3}(2860)\rightarrow
D^{*}K)}{\Gamma(D^{*}_{s3}(2860)\rightarrow DK)}= 0.38\\
R1 = \frac{\Gamma(D^{*}_{s3}(2860)\rightarrow
D_{s}\eta)}{\Gamma(D^{*}_{s3}(2860)\rightarrow DK)}= 0.03
\end{gather}
These ratios are compared with previous valued predicted by various
theoretical models as in Ref[5] $\Gamma(\frac{D^{*}K}{DK})= 0.39$
and $\Gamma(\frac{D_{s}\eta}{DK})= 0.13$ , Ref[10]
gives$\Gamma(\frac{D^{*}K}{DK})=0.43$ and
$\Gamma(\frac{D_{s}\eta}{DK})=0.11$ and Ref[7] gives
$\Gamma(\frac{D^{*}K}{DK})= 0.8$ and $\Gamma(\frac{D_{s}\eta}{DK})=
0.05$

\subsection{$1D_{s2}$ and $1D^{'}_{s2}$}

$1D_{s2}$ is the spin partner of the $D^{*}_{s1}(2860)$ belonging to
$J^{P}$ as $2^{-}_{\frac{3}{2}}$ state, and $1D^{'}_{s2}$ state
belongs to $J^{P}_{s_{l}}$ to $2^{-}_{\frac{5}{2}}$. These both
states are still unknown in the charm meson spectrum.

As shown in Table\ref{tab1}, their masses have been already
predicted by various theoretical models [21,22,23]. Taking their
masses to be within the allowed range 2800 MeV to 3000 MeV,
variation of their total OZI allowed two body strong decay width
have been plotted with respect to mass and coupling constant in
Figure8 and Figure9.

%\begin{figure}[h!]
%\includegraphics[width = 2in]{d2x}
%\includegraphics[width = 2.2in]{d2y}
%\caption{shows the variation of decay widths of $1D_{s2}$ and
%$1D^{'}_{s2}$ with their masses and their corresponding coupling
%constants.}
%\end{figure}
\begin{figure}[ht]
  \begin{minipage}[b]{0.5\linewidth}
    \caption{\small{Variation of $1D_{s2}$ with its mass and coupling constant
    gxh}}
    \label{ fig1:a)}
  \end{minipage}
  \begin{minipage}[b]{0.5\linewidth}
    \caption{\small{Variation of $1D^{'}_{s2}$ with its mass and coupling
    constnat}}
    \label{ fig1:b}
  \end{minipage}
  \end{figure}
Using the hadronic couplings obtained in previous subsections
$gxh=0.20$ and $gyh=0.40$ partial and total decay widths of these
states are listed in first column of Table\ref{tab5}. Also, these
two states can mix through spin-orbit interaction or by some other
mechanism and physically $D_{s2}^{'}$ and $D_{s2}$ can be
represented as the linear combination of $^{3}D_{2}$ and $^{1}D_{2}$
states as

\begin{center}
$ \left(
\begin{array}{c}
  1D(2^{-}) \\
  1D^{'}(2^{-}) \\
\end{array}
\right) = \left(
\begin{array}{cc}
Cos\theta_{1D} & Sin\theta_{1D} \\
  -Sin\theta_{1D} & Cos\theta_{1D} \\
\end{array}
\right) \left(
\begin{array}{c}
   1^{3}D_{2} \\
  1^{1}D_{2} \\
\end{array}
\right) $
\end{center}

where $\theta_{1D}$ is the mixing angle. In general the mixing angle
between the $^{3}L_{l}$ and $^{1}L_{l}$ in heavy quark limit is
given by $\theta_{1D} = arctan\sqrt{L/(L+1)}$. For this case the
mixing angle corresponds to $L=2$ and comes out to be $39.2^{\circ}
\sim 39^{\circ}$. In Table\ref{tab5}, the last column give partial
decay widths by taking this mixing into account.
% for various possible
%decay channels of these two states $1D_{2s}$ and $1D^{'}_{s2}$ have
%been listed with and without considering the effect of mixing angle.
\begin{table}
\begin{center}
\begin{tabular}{|l|r|r|r|r|}
\hline
 & \multicolumn{2}{c|}{As pure states}&\multicolumn{2}{c|}{As mixed states}\\
 &\multicolumn{2}{c|}{($\theta = 0^{\circ}$)}&\multicolumn{2}{c|}{($\theta = -39^{\circ}$)}\\
 \cline{2-3}\cline{4-5}
Decay Channel& $1 D_{2}$ &$1D^{'}_{2}$& $1 D_{2}$ & $1D^{'}_{2}$\\
&(MeV)&(MeV)&(MeV)&(MeV)\\

\hline
$D^{*}K$&61.37&20.51&36.02&56.95\\
 \hline
$D^{*}_{s}\eta$&16.69&2.51&11.62&13.42\\
 \hline
 $D(2400)K$&-&$7.5\times 10^{-5}$&-&$1.5\times 10^{-4}$\\
 \hline
 $D^{*}_{s}(2317)\eta$&0.0037&0.005&0.0014&0.01\\
 \hline
Total&78.06&21.04&47.65&70.38\\
\hline \hline
\end{tabular}

\caption{Calculated partial and total decay widths of $1D_{s2}$ and
$1D^{'}_{s2}$. First section is calculated by taking them as pure
states and the second section includes mixing scheme into account
}\label{tab5}
   \end{center}
\end{table}

 \section{Conclusion}

 Due to advancement in high energy accelerators, large amount of
 information is available on heavy-light charm and bottom
 mesons. These information motivates theorists to
 explore more about these heavy-light mesons. These D and B meson
 states are studied by observing their decaying behavior, masses, their
 $J^{P}$ states, coupling constants, branching ratios etc. Many
 models like Heavy quark effective theory, Quark pair creation
 model, Potential models etc are framed to study these heavy-light
 mesons. Recently, LHCb predicted spin 1 and spin 3 strange charm mesons. In this paper, we use the heavy quark effective approach to
 study the recently observed spin 1 and spin 3 strange charm
 states. This theory treat the heavy quark as static and provide lagrangian and decay widths formulas to the available states. This theory has adequately studied the previously determined
 experimental states and successfully allotted their positions in the
 charm and bottom spectroscopy.

 Observation of spin 1 and spin 3 resonances of $D^{*}_{s}(2860)$ by
 LHCb has clearly indicated that there are two different states of
 $D^{*}_{sJ}(2860)$. In the last 5 years, various theoretical models
 [5-7,10-19] studied $D^{*}_{sJ}(2860)$, favored it as $1 ^{3}D_{3}$ state
 with narrow decay width. From the LHCb data, $D^{*}_{s3}(2860)$
 state with $\Gamma=53 MeV$ can be correlated with this
 $D^{*}_{sJ}(2860)$ state. We too studied the decay behavior of
 $D^{*}_{s3}(2860)$ assuming it to be in the $1 ^{3}D_{3}$ state and
 calculated the hadronic coupling constant $gyh=0.40$. This value
 can be compared with the one obtained by Wang $gyh=0.52$ [25].

 We also studied the remaining spin 1 observed state by LHCb
 $D^{*}_{s1}(2860)$, assuming it to be pure $1^{3}D_{1s}$ state and to be a mixture of $1^{3}D_{1}$ and $2
 ^{3}S_{1}$ state. We study its decay channels ($D^{*}K,DK,D^{*}_{s}\eta,D_{s}\eta$) and  R value ($\frac{D^{*}K}{DK}$) calculated for the pure state (R=0.48)
  which does not lie
 within the given experimental data (R=1.10). So we adopted it to be as a
 mixture of radially excited $2 ^{3}S_{1}$ and orbitally excited $1 ^{3}D_{1}$. Using this interpretation, decay widths and R value depends on mixing angle $(\theta)$ and coupling constant $gxh$.
  We studied the variation of partial widths with coupling constant $gxh$ for some fixed values of mixing angle $(\theta)$ which shows $DK$ is the dominant decay channel. In the variation of R value with mixing angle $(\theta)$ ,
experimental R value favors the large mixing angle.
  This large mixing angle implies the predominance of $2
 ^{3}S_{1}$ state for $D^{*}_{s1}(2860)$. We obtained R=0.91 corresponding to the mixing
 angle $ -1.6 \leq \theta \leq -1.2$. Along with this mixing angle, we
 constrain the coupling constant $gxh$ to be lying in the range
 $ 0.17 \leq gxh \leq 0.20$. This obtained coupling value is close to the value given by Wang
$gxh=0.19$ in Ref[25].

Using these coupling constants, we also calculated the decay
behavior of the spin partners of these states $1 D_{2}$ and $1
D^{'}_{2s}$. These states are studied using two ways, first by
considering them as pure states and secondly by taking their mixing
into account. In both cases $D^{*}K$ is the dominating decay
channel. Decay width for $1 D^{'}_{2s}$ as a pure state comes to be
small indicating the presence of other decay modes. As we have only
considered the decays to pseudo-scalar mesons, so there may be a
possibility that decays to light vectors mesons may also be present
for this state.

\section{Acknowledgement}
The authors gratefully acknowledge the financial support by the
Department of Science and Technology (SB/FTP/PS-037/2014), New
Delhi.

\section{References}
\bibliography{latex-sample}
\bibliographystyle{unsrt}
\begin{enumerate}

 \item  B. Aubert, et al. (BaBar Collaboration), Phys. Rev. Lett.
97,222001 (2006).

 \item B. Aubert, et al. (BaBar Collaboration),
Phys. Rev. D 80,092003 (2009).

\item B. Zhang,X. Liu, W.Z. Deng and S.L. Zhu, Eur. Phys. J. C50(2007)
617.

\item P. Colangelo et al.,Phys. Rev. D
86,054024(2012)[arxiv:1207.6940/hep-ph]

\item P. Colangelo et al., Phys. Lett. B642 (2006)48.

\item  D. M. Li, B. Ma and Y. H. Liu, Eur. Phys. J. C51 (2007) 359,[hep-ph/0703278].

 \item D. M. Li, B. Ma, Phys. Rev. D81 (2010)014021.

\item R.Aaij et al. [LHCb collaboration],Phys. Rev. Lett. 113, 162001
(2014),arXiv:1407.7574 [hep-ex].

\item R.Aaij et al. [LHCb collaboration],Phys. Rev. D 90, 072003
(2014),arXiv:1407.7712 [hep-ex].

\item X. H. Zhong and Q.Zhao,Phys. Rev. D78 (2008) 014029.

\item X. H. Zhong and Q. Zhao,Phys.Rev. D81 (2010) 014031.

\item Zhi-Gang Wang,Eur.Phys.J.C75(2015)25, arxiv: 1408.6465v3[hep-ph]
(2014).

\item  J. Vijande, A.Valcarce and F. Fernandez, Phys.
   Rev.D79,037501 (2009)[arxiv:0810.4988[hep-ph]].

\item Stephen Godfrey and Ian T. Jardine,Phys. Rev. D 89, 074023 (2014), arxiv: 1312.6181v2.

\item Qing-Tao Song et al.,Eur. Phys. J. C75 (2015) 1,30.

\item Ling Yuan et al. arxiv:1203.0370v1 (2012).

\item F.E.Close et al., Phys. Lett. B647,159(2007).

\item De-Min Li et al., Eur. Phys.J.C71 (2011) 1582.

\item Bing Chen et al. "combined study of 2S and 1D open charm
   strange mesons",    Phys. Rev. D 92, 034005 (2015), arxiv:1507.02339v2.

\item K. A. Olive et al. (Particle Data Group),Chin. Phys. C, 38,
   090001 (2014).

\item S.Godfrey and N. Isgur, Phys. Rev. D32,189(1985).

\item M.Di Pierro and E. Eichten, Phys. Rev. D64 (2001) 114004
   [hep-ph/0104208].

\item D. Ebert, R. N. Faustov and V. O. Galkin, Eur. Phys. J. C 66
   (2010) 197 [arxiv:0910.5612 hep-ph]

\item Introduction to Heavy Quark effective Theory by A.G.Grozin.

\item Zhi-Gang Wang,Eur.Phys.J.Plus 129 (2014) 186, Meenakshi Batra et al. Eur.Phys.J. C75 (2015) 319.

\item P. Colangelo et al.,Mod.Phys.Lett. A19 (2004) 2083-2102.

\end{enumerate}

\end{document}